# Predicting respondent difficulty in web surveys: A machine-learning approach based on mouse movement features

Amanda Fernández-Fontelo[1*], Pascal J. Kieslich[2], Felix Henninger[2,3], Frauke Kreuter[2,4,5], Sonja Greven[1]

[1] Chair of Statistics, School of Business and Economics, Humboldt-Universität zu Berlin, Berlin, Germany

[2] Mannheim Centre for European Social Research, University of Mannheim, Mannheim, Germany

[3] Cognitive Psychology Lab, University of Koblenz-Landau, Landau, Germany

[4] University of Maryland, College Park, Maryland, USA

[5] Institute for Employment Research, Mannheim, Germany

*Corresponding author: Amanda Fernández-Fontelo, HU, Berlin

## Abstract

A central goal of survey research is to collect robust and reliable data from respondents. However, despite researchers' best efforts in designing questionnaires, respondents may experience difficulty understanding questions' intent and therefore may struggle to respond appropriately. If it were possible to detect such difficulty, this knowledge could be used to inform real-time interventions through responsive questionnaire design, or to indicate and correct measurement error after the fact. Previous research in the context of web surveys has used paradata, specifically response times, to detect difficulties and to help improve user experience and data quality. However, richer data sources are now available, in the form of the movements respondents make with the mouse, as an additional and far more detailed indicator for the respondent-survey interaction. This paper uses machine learning techniques to explore the predictive value of mouse-tracking data with regard to respondents' difficulty.

We use data from a survey on respondents' employment history and demographic information, in which we experimentally manipulate the difficulty of several questions. Using features derived from the cursor movements, we predict whether respondents answered the easy or difficult version of a question, using and comparing several state-of-the-art supervised learning methods. In addition, we develop a personalization method that adjusts for respondents' baseline mouse behavior and evaluate its performance. For all three manipulated survey questions, we find that including the full set of mouse movement features improved prediction performance over response-time-only models in nested cross-validation. Accounting for individual differences in mouse movements led to further improvements.

**Keywords:** mouse movements, paradata, web surveys, difficulty, personalization, supervised learning models, classification

# Introduction and background

Two decades ago Mick Couper coined the term paradata in a presentation at the Joint Statistical Meetings (1998) to describe data that is (ideally) an automated by-product of the survey data collection, or that can be collected alongside the data collection process. He encouraged data collectors to make systematic use of such by-products to learn about the collection process and ideally improve it. The message took and survey organizations have vastly increased their use of paradata (Kreuter, 2013; McClain et al., 2019). Most collections and applications of paradata are to monitor fieldwork efficiency (Vandenplas et al., 2017), monitor interviewer behavior (Sharma, 2019), or to improve nonresponse adjustment (Olson, 2012). Increasingly we see applications in adaptive survey designs (Chun et al., 2017), though these affect mostly the allocation of fieldwork resources, and not a paradata driven adaptation of questionnaires (Callegaro, 2013; Early, 2017). In the current contribution, we focused on a particular type of paradata – participants' cursor movements during a survey – and examined if they could be used to predict response difficulty in online surveys.

In light of increasing survey costs and decreasing response rates, online surveys have become very prominent across many fields and data collection settings (Couper, 2011), with even massive governmental data collection efforts moving to the web (U.S. Census Bureau, 2017). This shift in medium poses some unique challenges, though it shares with other survey modalities the risk that – despite the careful design and testing of questionnaires (Presser et al., 2004) – respondents regularly experience difficulty in understanding what the question is asking (Tourangeau et al., 2000) or how the question's concepts apply to respondents' circumstances (Conrad & Schober, 2000; Ehlen et al., 2007; Schober et al., 2004), and may provide incorrect responses as a result. In contrast to web surveys, however, other modalities provide ways of mitigating the risk of misunderstandings: In face-to-face and telephone interviews, for example, interviewers can give and pick up paralinguistic information (de Leeuw, 2005; Tourangeau et al., 2013), and help respondents accordingly. Self-administered web surveys provide no such interaction, leaving respondents to their own devices.

One hope, expressed by survey methodologists (see for example Callegaro, 2013), is that paradata might be used to detect respondents facing difficulties as they complete online surveys: If paradata could pick up signals for struggling participants (just like an interviewer would), web surveys could 'react' and offer help or, in the analysis, respondents' answers could be treated with caution when strong indications for misunderstandings are present. Prior research has considered multiple distinct sources of paradata, which we describe and discuss in the following.

So far, *response times* have received the most attention in paradata research. The notion that response times – and long idle times in particular – are indicative of difficulty processing a question is not new. For example, website design already makes use of information on idle periods, automatically triggering virtual assistance (Conrad et al., 2007). Mittereder (2019) used response time to predict break-offs in web surveys and design interventions.

Keystrokes and in particular response times have also been examined in their relationship to measurement error (Heerwegh, 2003). Most work so far has focused on the time between reading questions and providing a final answer, some also in combination with changes in the response (Heerwegh, 2003; Yan & Tourangeau, 2008; Stern, 2008; Zhang & Conrad, 2014). Very low and very high response times are often used as indication for bad data quality (Conrad et al., 2006, 2007; Yan et al., 2015; Yan & Olson, 2013). However, there are several limitations to using response times outside of a laboratory environment. First, response time is a relatively coarse measure that does not specify what might have caused any observed latency: Respondents who take a long time to answer may not even be engaged in the survey task, but checking email, talking on the phone or tending to a child (Höhne & Schlosser, 2018; Sendelbah et al., 2016). Also, direct comparisons of absolute response times between participants may be problematic because each person has his or her own baseline speed (Mayerl et al., 2005), and models for response times that include task characteristics and respondent features are still in their infancy (Couper & Kreuter, 2013).

*Mouse movements* provide a promising source of information regarding the cognitive processes underlying choices. In the cognitive sciences, features collected via mouse-tracking are frequently used to assess how the commitment to choice alternatives develops over time and to quantify the amount of conflict participants experience while making their decision (see recent reviews by Freeman, 2018; Stillman et al., 2018). Most mouse-tracking studies have focused on experimental manipulations to test theoretical predictions regarding how characteristics of the decision situation and the respondent influence decision conflict (see Freeman & Ambady, 2011, for an early review). Several more recent studies, however, used mouse-tracking features to predict decisions and demonstrated that they can add additional information beyond response times. For example, in an intertemporal choice task, features extracted from early mouse movements predicted the subjective value of participants' later choices – independently of participants' response times (O'Hora et al., 2016). In another study, participants' average conflict in a self-control task (as assessed through the curvature of their mouse movements) but not their response time predicted their decision between healthy or unhealthy food at the end of the experiment (Stillman et al., 2017). However, given that these findings were obtained in laboratory tasks with a vastly simplified, artificial screen layout, it remains an open question if they generalize to survey research.

Previous research also indicated that mouse-tracking could provide a useful data source in a questionnaire context: Stieger and Reips (2010), for example, showed that excessive mouse movements (as defined by distance) identify respondents with low data quality. In a laboratory study, Horwitz et al. (2017) classified specific mouse movement patterns through manual coding and demonstrated that they are useful additions to response time when predicting response difficulties. Specifically, they identified periods during which the cursor paused on top of the question text (hovers) or a response option (markers) for two or more seconds, as well as regressive movements between different areas of the page. However, any larger-scale application in online surveys should automate the processing of mouse-tracking data and the computation of mouse movement features. Concerning the automatic computation of measures, the cognitive sciences literature provides a wealth of quantitative mouse-tracking measures that capture different aspects of the response process, such as the speed of the cursor movement, the number of changes in direction, and periods without movement (discussed in more detail below, see also Horwitz et al., 2019; Kieslich et al., 2019). Together, these indices provide a comprehensive picture of the response process, while being computable efficiently and in real-time, making them applicable even in large surveys outside of a laboratory setting.

In our study, we went beyond the prior literature in multiple ways: Compared to the study by Horwitz et al. (2017), we automatically extracted a set of commonly used quantitative mouse-tracking measures rather than coding them manually. In addition, we left the controlled environment of the laboratory and used a large-scale dataset collected in an online survey. Therein, we experimentally manipulated difficulty in a set of questions (described below in more detail). We reported an initial analysis of this dataset in Horwitz et al. (2019), showing that many indices, when analyzed in isolation, were affected by the difficulty manipulations. By contrast, our focus in the current contribution was on the predictive power of paradata, which is a critical prerequisite for their practical use. Thus, we examined not only if mouse movements differ between questions according to difficulty, but whether they provided sufficient predictive performance to recover the experimental group, and which features of mouse movements were particularly important.

Building on the data from Horwitz et al. (2019), we further investigated whether using multiple indices could jointly improve predictive accuracy further. Notably, the presence of a significant difference in certain mouse movements between easy and difficult versions of a question is, while an indicator, no guarantee for the predictive power of a specific feature (Lo et al., 2015; Shmueli, 2010). Possible reasons include that significance does not necessarily imply large effect sizes, goodness-of-fit in-sample does not guarantee predictive accuracy out-of-sample (Yarkoni & Westfall, 2017) and that significant variables may show an association with the outcome only in a small subgroup, leading to poor population-wide prediction (Lo et al., 2015). Therefore, we investigated whether the added predictive power of mouse movement features was sufficiently large for their use in informing real-time responsive questionnaire design or in measurement error corrections.

Finally, we also examined whether accounting for individual differences could aid the prediction of difficulty: Whether due to habit or preference, differences in hardware or system settings, the interaction with the survey, and, as a result, mouse movements, may vary systematically between respondents (Henninger & Kieslich, 2020). It is thus likely that focusing on deviations relative to a previously observed baseline (e.g. 'unusual for this subject' behavior) rather than an absolute value will reduce interindividual variation present in the data, and further strengthen predictive performance. However, this remains to be shown empirically, and thus we examined the effect of personalized predictive models in our analysis.

To summarize, we examined the predictive value of mouse movements as indicators of difficulty in a survey context. Our main questions were: When predicting response difficulty, what do we gain beyond response time through mouse-tracking features, and can we further improve prediction through personalization? Practitioners will want to know which machine learning algorithm should be used, and we will discuss this in the paper as well.

# Data and Methods

## Survey data description

The analyses in this paper were based on a survey conducted for the Institute for Employment Research (IAB) in Nuremberg, Germany, from September to October 2016 (Horwitz et al., 2019). The survey contained questions on a range of topics, with a focus on the respondents' employment history and demographic information. Recruitment was based on a set of 1627 respondents who had participated in a previous wave and agreed to future contact, and received an email or a postal invitation if an address was on file; 1527 individuals were also given a 5€ incentive (while the first 100 individuals recruited via email did not receive an incentive for participation). Data collection took place entirely online through a web survey (constructed in SoSciSurvey; Leiner, 2014). In total, 1250 participants responded, and 1213 completed the questionnaire. Of these, 886 (73%) reported using a mouse as an input device; our analysis is limited to these participants. The average age of these participants was 51 years (*SD*=10.8 years) and there were 454 female and 425 male participants (2 selected the other category, 5 did not provide an answer). We applied a number of additional, question-specific exclusions which are discussed below.

Our current analyses focus on a subset of questions concerning respondents' employment history and demographic information, all based on a multiple-choice format (screenshots of all relevant questions are provided in the Online Supplementary Material). Three questions were the focus of the following analyses (target questions): One assessed respondents' type of employment (*employment detail* in the following; e.g., employed at a private, profit-oriented company or in the civil service of a federal state); another the employees' position in the company hierarchy (*employee level*, e.g., executing occupation following instructions or comprehensive decision-making powers); the remaining target question assessed participants' highest attained level of education in the German school system (*education level*). For this question alone, open text inputs allowed further specification of the chosen option. A set of eight additional questions provided a baseline for participants' interaction behavior with the survey, using a similar format to the target questions but excluding any difficulty manipulation. They covered participants' evaluations of the general and personal current economic situation, their employment contract type, current job position, satisfaction with their employee representatives, if they were working in marginal employment, if they received unemployment compensation in the last year, and whether their salary increased enough to make up for inflation.

For each of the three target questions, every participant was randomly assigned to one of two difficulty levels designed to make responding more or less difficult (see Horwitz et al., 2019). The survey literature has discussed a number of factors that influence how easily participants understand and can answer a question. Among other things, these include aspects of the question wording and the response format (Holbrook et al., 2006; Lenzner et al., 2010). For employment detail, we therefore manipulated the wording of the response options, which was either *straightforward* with concise and simple vocabulary and grammar, or involved longer and more *complex* descriptions and writing structure, which should make understanding and answering the question more difficult. In addition, we manipulated the order of the response options for the employee and education level questions, with *ordered* response options in one condition (i.e., increasing from low to high levels) and randomly *unordered* options in the other. We implemented a balanced assignment independently for each question, and their ability to recover the experimental condition based on mouse movements will serve as the criterion of the predictive models in the following.

## Mouse movement trajectories

For the entire duration of the survey, paradata were gathered using a client-side collection script (Henninger & Kieslich, 2020) and transferred to the server in ten-second increments. As a preprocessing step, we extracted the trajectories from the paradata and applied a number of filtering operations to ensure a consistent dataset for each target question. First, participants who did not provide an answer (either because the question was not presented to them, e.g., if they were not an employee for the employee level question, or simply because participants did not select an answer) were excluded. Next, we excluded questions for which mouse movements were not recorded or incomplete (e.g. because of intermittent connection issues) and those for which paradata indicated that participants might have reloaded the survey page. For the education question, we also excluded participants who had selected one of the further options with a free-form text input field as opposed to a predefined choice option. As all models control for participants' gender and age, we excluded participants with missing values on these questions (as well as participants who selected the "other" category for gender, since there were too few observations in this category to include it as a predictor). As a final criterion, we removed instances in which participants took unrealistically long to answer a particular question (response time > 7 minutes). Applying all filter criteria resulted in a final data set of 551 participants for the employment detail question, 501 participants for employee level, and 548 for education level.

[Insert Table 1 about here]

From the recorded trajectories, we calculated a variety of mouse-tracking indices common in and adapted from the psychological process tracing literature (see also Freeman & Ambady, 2011; Kieslich et al., 2019) and the survey literature (see Horwitz et al., 2017) to capture distinct features of cursor trajectories on every page. These indices are described in Table 1. The processing of the collected mouse-tracking data and the calculation of the described indices was automated through the mousetrap package for the statistical programming language R (Kieslich et al., 2019), which provides extensive tools for processing and manipulating trajectory data. Mouse-tracking measures as aggregates of the far more complex cursor movements greatly reduce the amount of data, and while we computed indices post-hoc based on the raw trajectory records, the same analyses could, in the future, potentially be conducted in real-time on the client side.

Based on the survey literature summarized above, we hypothesized that several paradata indices could indicate response difficulty, including prolonged response times (Conrad et al., 2007; Mittereder, 2019), longer distances traveled (Stieger & Reips, 2010), and a greater number of hovers and y-flips (Horwitz et al., 2017). However, mouse-tracking applications in laboratory studies in the cognitives sciences suggest that other measures, including initiation times and indices concerning velocity and acceleration, are related to response competition and hence play a role for detecting response difficulty as well (see overviews by Hehman et al., 2015; Stillman et al., 2018). However, it remains an open question which of these indices predicts response difficulty in a joint analysis, and how well all indices together are able to detect response difficulty.

**Classification supervised learning methods**

To predict difficulty from mouse movements, we used different supervised classification methods. These map a categorical response variable (output or target) to explanatory variables of any type (inputs or predictors) through a specific function (e.g., logit). Our target variable was binary (difficult vs. easy) and we had eleven explanatory variables (mouse-tracking measures as described in Table 1, as well as age and gender) for each question. Each model is fit on a training sample, and its predictive performance evaluated on the remainder of the dataset (Hastie et al., 2009; James et al., 2013). To cover different relationships between the outcome and inputs, we considered the following supervised predictive models: logistic regression, tree-based models (classification trees, random forest, and boosting), support vector machines, and single hidden layer back-propagation networks.

*Logistic regression* models assume a linear relationship between the log-odds of the binary outcome Y (here: $Y = 0$ non-difficulty, and $Y = 0$ difficulty), and the vector of predictors $X = (X_1, X_2, \ldots, X_p)$ through the following equation:

$$\log(odds(Y = 1|X)) = \log\left(\frac{P(Y = 1|X)}{1 - P(Y = 1|X)}\right) = \beta_0 + \beta_1 X_1 + \beta_2 X_2 + \cdots + \beta_p X_p \quad (1)$$

Where $\log\left(\frac{P(Y=1|X)}{1-P(Y=1|X)}\right)$ is the logistic function of $P(Y=1)$ and acts to linearize the relationship between the output and inputs. The model parameters $\beta_1, \ldots, \beta_p$ quantify the relationship between the output and the corresponding input (see Hosmer et al. 2013).

*Classification trees* partition the space of the vector of predictors into a set of different subspaces. This division process is recursively executed and consists of splitting each predictor at a collection of possible cut-points and selecting the partition that most reduces the impurity within nodes (regions defined by the splits), as measured through the Gini index. To control overfitting and reduce tree complexity, trees can be pruned, trading a small increase in bias for a larger reduction in variance of predictions. These methods, however, can be unstable in that small variations in the training sample can change the tree structure dramatically (Breiman et al., 1984; Loh, 2014).

Bootstrap aggregating (bagging) mitigates this issue using bootstrap routines (Efron, 1979; Efron & Tibshirani, 1993) and aggregating, which decreases the variance of the predictions, while also increasing the model's accuracy. Bootstrap samples are drawn from the training data and an unpruned tree fit to each to ensure little bias. Bootstrap trees are aggregated for an overall prediction (e.g., using the majority vote for class membership). As a high correlation between bootstrap trees diminishes the possible variance reduction in bagging, *random forests* consider a new random subsample of m predictors for each bootstrap tree. This includes weaker predictors in some of the bootstrap trees, increasing variability and decreasing correlation, reducing the variance of the overall prediction. The tuning parameters are the number of bootstrap trees and the number of selected predictors m. See Breiman (1996, 2001), Dietterich (2000), and Sutton (2005).

Gradient boosting methods were also designed to reduce the trees' variance and improve the models' performance. Here, the weak classifiers (e.g., trees) are grown sequentially. Boosting iteratively fits a new tree based on the information that was not explained in the previously grown trees (so-called pseudo-residuals, serving as responses in the next iteration). The main gradient boosting tuning parameters are the number of trees, the number of splits in each tree, and a shrinkage parameter that controls how the boosting model learns. See Breiman (1998), Friedman (2001), and Schapire (2003).

Support vector machines are used to construct non-linear decision boundaries to discriminate between different categories. A linear boundary can be defined as the hyperplane in the covariate space that gives the largest margin (distance) to the nearest observations (support vectors) in the two classes on both sides of the plane (support vector classifier), while allowing for some misclassifications. As samples are usually not linearly separable into two classes, kernel functions (here the radial kernel was considered) enlarge the predictors' vector space and assure that the decision boundary is no longer linear in the original predictors' vector space. The two tuning parameters are a regularization and a kernel-related parameter. See Boser et al. (1992), Cortes and Vapnik (1995), and Steinwart and Christmann (2008).

Single hidden layer back-propagation networks consist of inputs, a single hidden layer with h nodes, and outputs that can be represented as a graph. Each node of the hidden layer corresponds to an activation function evaluated at a linear combination of the inputs, where parameters act as weights. The outputs are also non-linear transformations (e.g., logistic transformations) of linear combinations of these nodes in the hidden layer. The size of the hidden layer can be tuned by constructive or deconstructive approaches. They start with a minimal or oversized network and grow or prune that network (e.g. using cross-validation). In addition, a regularization parameter (weight decay) is usually considered in the loss function to prevent overfitting. See Ripley (1994a, 1994b).

For a detailed summary of the methods described in this section see also Hastie et al. (2009), and James et al. (2013).

## Model tuning of parameters, performance and importance measures

We used cross-validation as one of the most common methods for both tuning parameters and evaluating a model's predictive performance. In cross-validation, the sample is divided into K parts with each part left out in turn as a testing set for the model fit on the training set (the remaining K-1 parts). Parameters are chosen to optimize the average prediction error over the K left-out parts. As the parameters are tuned to a given data set and to get an honest assessment of the out-of-sample prediction error, we used nested cross-validation. Here, for each left-out part of the data (validation set) in the outer loop, cross-validation with further splits of the remaining data into training and test sets is performed (inner loop). While the tuning parameters for each outer fold are chosen using the inner loop, out-of-sample prediction error of the tuned models is evaluated on the validation sets in the outer loop. This avoids over-optimistic performance evaluation when tuning and validating on the same samples (Varma & Simon, 2006). For more details, see Efron and Tibshirani (1997), Hastie et al. (2009), and James et al. (2013).

We used nested cross-validation for all tuning parameters with 10-fold cross-validation for the outer split and subsampling (500 repetitions) with weights of 75% and 25% for the sub-training and test sets in the inner splits. Since the outcomes are balanced (Wei & Dunbrack, 2013), we used the proportion of correctly classified observations (accuracy) to evaluate performance in both inner and outer loop.

All supervised learning models were computed with the mlr package in R 3.6.1 (Bischl et al., 2020) using parallelization with 32 CPUs. Equal training and testing samples were used across the different models for each question to ensure comparability. Results in Table 2 are reproducible if the same seeds and the same number of CPUs are used.

Permutational feature importance measures (Strobl et al., 2008) were computed to extract the effect of each measure in the models. These permute the values of a measure randomly and assess how the overall prediction of the given model is affected. We generated 500 independent permutations for each measure and assessed the models' prediction performance through the accuracy.

**Personalization**

We investigated personalization of the measures to reduce variability not related to the classification task and to correct for different baseline behavior, making the measures more comparable between respondents. Measures may additionally be influenced by characteristics of the question such as the question layout or the answer given. For example, response time or total distance traveled may be larger if the position of the answer is further away from the submit button, but this is not related to the question difficulty.

We proposed two methods of personalization: one that only corrects for the baseline behavior of the respondents using the eight non-manipulated baseline questions, and a second that corrects for the baseline behavior and the position of the answer. We regressed every mouse-tracking measure onto its values in all baseline questions, and on a position indicator for the second correction method, and performed our analyses using the residual values. Specifically, to correct for the baseline behavior of the respondents, denote by $B_{ijr}$ the j-th measure (e.g., response time) of the i-th individual in the r-th baseline question (e.g., general economy). Let $Y_{ijk}$ be the j-th measure of the i-th individual in the k-th target question (e.g., employment detail). Consider the following multiple linear regression model:

$$Y_{ijk} = \alpha_{j0} + \sum_{r=1}^{8} \alpha_{jr} B_{ijr} + \epsilon_{ijk} \qquad (2)$$

where $\alpha_{jr}$ quantifies the association between the j-th measure in the r-th baseline and k-th target questions respectively, and $\epsilon_{ijk}$ is the error or the part of $Y_{ijk}$ not explained by baseline behavior. We used the model residuals $\hat{\epsilon}_{ijk}$ as baseline-corrected measures.

To additionally remove the effect of the position of the responses, we used a two-step method. In the first step, each measure of each target question is corrected for the corresponding positions with the linear regression

$$Y_{ijk} = \gamma_{jk0} + \sum_{m=1}^{m_k-1} \gamma_{jkm} P_{ik}(m) + w_{ijk} \qquad (3)$$

Where $P_{ik}(m)$ is an indicator whether the i-th individual's answer in the k-th target question was at the m-th position (out of $m_k$ options). The residuals from (3), $\widehat{w}_{ijk}$, correspond to variability of the j-th measure in the k-th target variable not explained by the response position. Corrected measures $\widehat{w}_{ijr}$ for the baseline questions are constructed in the same way.

The second step additionally corrects for the individual characteristics using the simple linear regression:

$$\widehat{w}_{ijk} = \phi_{j0} + \phi_{j1}\overline{\widehat{w}_{ij}} + v_{ijk} \qquad (4)$$

where $\overline{\widehat{w}_{ij}} = \frac{1}{8}\sum_{r=1}^{8} \widehat{w}_{ijr}$ averages over the residuals from (3) within the same individual and the same measure over the eight baseline questions. Finally, the residuals $\widehat{v_{ijk}}$ are the position- and baseline-corrected measures.

# Results

Employment detail had nine possible answer options. Of the 551 participants that were included in the analysis, 49.5%, 17.1%, and 7.1% indicated that they were employed in the civil service at the municipality, federal state and national levels, respectively. In addition, 22.0% and 4.4% were employed at profit-oriented and nonprofit-oriented private companies, respectively. Due to the filter criteria applied above, there were no participants that had selected the other alternatives (self-employed independent, self-employed not independent, freelance, and assisting family members).

Employee level had only four answer choices. Of the 501 participants that were included in the analysis, 8.6% answered that they had an executing occupation following instructions (e.g., secretarial or nursing assistant) and 47.7% had a qualified occupation following instructions (e.g., accountant). Also, 40.7% responded that they had an occupation with some independent activities and responsibilities (e.g., scientific employee, department manager), and 3.0% indicated they had comprehensive management tasks (e.g., director or member of the executive board).

Education level had eleven answer options. Of the 548 participants that were included in the analysis, 35.0% had a general qualification for university entrance and 14.2% a qualification for universities of applied sciences. 27.2% had completed higher secondary education (Realschulabschluss) and 10.0% lower secondary education (Hauptschulabschluss). The rest had graduated from Polytechnic High School in 8th grade (0.4%) or 10th grade (7.1%) (DDR graduations). 5.8% completed Extended Secondary School (DDR graduation) and 0.2% indicated no graduation.

Each of the measures in Table 1 extracted from the mouse movement trajectories may capture different details of the response process, which might differ between difficult and non-difficult settings. Figure 1 shows the empirical distribution of these measures for both difficult and non-difficult settings for employment detail, employee level and education level respectively. The same figures for baseline-corrected, and baseline and position-corrected measures for the three target variables can be found in the Supplementary Material.

For each target question, each supervised predictive model described in Data and Methods was considered as a potential classifier using the experimental condition (difficult or easy) as a target variable, and the measures in Table 1, and age and gender as predictors. When personalization was also considered, the measures in Table 1 were corrected before being included as predictors.

[Insert Figure 1 about here]

Results in Table 2 show the best models, in terms of accuracy, for each question and each personalization approach. Additionally, there were two different types of models: full and response-time-only models to investigate the gain in accuracy or added value of the extracted mouse tracking measures over just response time for different kinds of difficulty. Both the full model with all nine mouse movement measures and the response-time-only version also included age and gender as predictors. More details on the full results for each model and question, and the corresponding R codes are included in the Supplementary Material. Hover-type measures are computed depending on a threshold that has to be chosen, usually empirically. For instance, in Horwitz et al. (2017), the authors considered 2000 ms as a threshold for hovers. To investigate dependence on this parameter, we conducted an extensive study considering this set of potential thresholds: 250 ms, 500 ms, 2000 ms, and 3000 ms. As shown in Table 2, no threshold was uniquely optimal for all questions and results were similar across thresholds in many cases.

For employment detail, gradient boosting with age, gender and nine baseline-corrected measures (full model) performed best among the predictive learning approaches. On average, this model correctly classified 65.9% of the observations; in particular, the model classified 56.3% of the non-difficult scenarios correctly (concise and straightforward language), and 74.2% of the difficult scenarios (complex language). The best full model with uncorrected measures only provided an accuracy of 61.0%, indicating the necessity for personalization. The best response-time-only model showed an accuracy of 64.8%, a bit lower than the accuracy of the best model, indicating a small gain of using all mouse tracking measures over the response-time-only model. Figure 2 shows the permutational importance features (Strobl et al., 2008) of the measures in the gradient boosting model. The green points in the graph describe the differences between the mean accuracies of gradient boosting with the observed measures and with the measures randomly permuted one by one. The more negative the difference, the more impact the corresponding measure has on the model's predictive performance. Accordingly, the most important measures in this gradient boosting model were response time, and the vertical and horizontal flips with an average decrease in the model's accuracy of 0.142, 0.028 and 0.014, respectively.

For employee level, gradient boosting with all baseline- and position-corrected measures as predictors again performed best in terms of accuracy. In particular, 59.1% of the observations were correctly classified; 52.6% and 65.3% of the non-difficult and of the difficult scenarios were classified correctly, respectively. The best full model for the uncorrected measures and the best response-time-only model showed accuracies of 55.5% and 55.7%, respectively, both again smaller than that of the overall best model. Figure 2 shows that the most important measures in the gradient boosting model for employee level were initiation time, horizontal flips, and hovers with an average decrease in the model's overall accuracy of 0.113, 0.041, and 0.033, respectively. However, for this particular question, permuting response time decreased the overall accuracy by only 0.009, indicating the necessity of more measures to predict the difficulty in this ordered vs. unordered manipulation.

For education level, a random forest with eleven baseline- and position-corrected measures performed best in terms of accuracy. This model correctly classified 58.9% of the observations, with non-difficult and difficult scenarios correctly classified with rates of 62.3% and 55.12%, respectively. The best full model for the uncorrected measures and the best response-time-only model only showed accuracies of 56.2% and 56.4%, respectively. Figure 2 shows that the most important measures in the random forest were maximum acceleration, initiation time, and hovers with an average accuracy decrease when permuted of 0.224, 0.049, and 0.033, respectively. For this question again, randomly permuting response time did not decrease the overall accuracy notably.

[Insert Table 2 and Table 3 about here]

[Insert Figure 2 about here]

Overall, we found that for all three target questions, inclusion of all mouse movement measures improved predictive accuracy compared to the response-time-only model, indicating that mouse movements contain more information about respondents' difficulty to answer questions than the response time alone. An even stronger gain in accuracy was evident when personalizing the mouse movement features, indicating the importance of taking different behavior of respondents with the mouse into account. Taking the position of the answer category into account was additionally beneficial if the answers were differently ordered. We also found that while there were clearly significant differences in mouse trajectories (as measured by the extracted indices) between difficult and non-difficult settings, the manipulations were not strong enough to allow for reliable prediction of difficulty from the mouse movement indices alone. Finally, we saw that tree-based learning models (decision trees, random forest, and gradient boosting) are those that usually predicted the difficulty best across questions. In particular, Table 3 shows that, for both employee and education levels, the best accuracies and the accuracies of the other tree-based models differed by roughly 0.012 and 0.015, respectively, with specificity and sensitivity values being roughly balanced. More substantial raw differences of 0.12, 0.02 and 0.07, and 0.11, 0.13 and 0.07 were observed between the best accuracies and the accuracies of the logit, support vector machine, and neural network models for employee and education level, respectively. For employment detail, raw differences between the best accuracies and the accuracies of tree- and non-tree-based models were generally smaller, except for the worse performing neural networks. However, the performance of tree-based models was still satisfactory and included the best model.

# Discussion

This work aimed at predicting respondents' difficulty in a web survey with a set of mouse movements features commonly used in cognitive science and the survey literature. For prediction, we applied and compared a set of different machine learning methods. We found that the use of several mouse movement measures improved the prediction of respondents' difficulty, above and beyond the use of response times. We also saw that further improvements in prediction were achievable by controlling for between-participant differences in the measures with baseline questions.

Respondent difficulty is one important cause of measurement error in web surveys and can lead to poor data quality and potentially weaken or bias results and conclusions. The detection of such difficulty is an important step in developing corrections for measurement error when analyzing survey responses and could potentially even allow for implementing real-time interventions while participants fill out a survey. Real-time interventions could reach from pop-up help screens, reminders to respond carefully, all the way to chat assistance, either by a bot or a human. In order to both not miss a respondent experiencing difficulty while at the same time not to bother a respondent unnecessarily with such interventions (and increase response burden because of this), the good performance of any predictive model triggering the intervention is key.

For the three target questions analyzed here, we found that the best predictive models were tree-based models (classification trees, random forest, and gradient boosting) that use baseline- or, if the position differs by experimental conditions within the same questions (e.g., ordered vs. unordered answering options), position- and baseline-corrected measures, and that response time was not always the most important measure for predicting difficulty. Specifically, response time was the most relevant measure in the best predictive model for employment detail (simple vs. complex language). However, measures such as initiation time, maximum acceleration, horizontal flips, and hovers improved difficulty prediction more than response time for both employee and education levels (ordered vs. randomly ordered response options). This might indicate that different manipulations produced different types of difficulty, which can be captured through different paradata sources. In addition, the most important features differed by question even if these were manipulated in the same manner and difficulty was potentially similar. For example, maximum acceleration was the most important measure for education level but not useful for employee level difficulty prediction. This could be explained by the fact that the number of response options was considerably larger for the education level question and that participants in the unordered condition for this question might have more often and more strongly showed variations in their cursor speed. Even when using a large set of mouse-tracking measures and accounting for individual variability, there is still considerable room for improvement when predicting response difficulty. Some likely explanations could be related to the intensity of the manipulations. Even though response difficulty was experimentally manipulated to create two controlled scenarios of difficulty, manipulations changed only a very specific aspect of difficulty each. Hence, the average degree of response difficulty might not have varied very strongly between conditions; however, given that we only manipulated but did not measure difficulty we cannot quantify this in the current study. Also, the strength of these manipulations may not have been comparable for all participants and questions and we might thus observe a mixture of behaviors. In addition, since the different difficulty manipulations were varied between questions, we cannot disentangle effects of the specific question from effects of the type of difficulty (complex language vs. random ordering of answer options), e.g. on the most relevant feature. A complementary approach could in a future study measure participants' subjective difficulty for a given question and use it as the outcome for prediction (Horwitz et al., 2017). Future research could also include additional difficulty manipulations and manipulate different types of difficulty within the same question in a crossed design.

Although mouse movement measures have shown potential for difficulty prediction, they use only summaries of the information contained in the respondents' cursor movements. The use of full mouse movement trajectories as bivariate functions may further improve prediction results compared to the extracted mouse movement features, if suitable functional data methods are developed. In addition, the information from mouse movements could in a future study be enriched by additional information such as respondents' click data and changes in the response options. In further research, we also plan to use linked administrative data that allows us to quantify measurement error in certain survey responses and investigate an analogous prediction of measurement error.

## Acknowledgements

The authors would like to thank the Institute for Employment Research (IAB) for their support in conducting the survey that provided the data for the current analyses, and in particular Malte Schierholz and Ursula Jaenichen for providing guidance and insights into the data. In addition, the authors would like to thank student assistants Anja Humbs, Zhang Ran and Franziska Leipold for their help in setting up the original survey, running the analyses and with the reference management.

## Funding

The authors acknowledge financial support from the German Research Foundation (DFG) through the grant "Statistical modeling using mouse movements to model measurement error and improve data quality in web surveys" (GR 3793/2-1 and KR 2211/5-1).

## Author information

**Amanda Fernández-Fontelo** is a postdoctoral researcher at the Chair of Statistics at the Humboldt-Universität zu Berlin, Germany. She earned her master's degree in Statistics, Software and Operations Research at the Universitat Politècnica de Catalunya in 2015, and her Ph.D. in Mathematics at the Universitat Autònoma de Barcelona in 2018. Her main research is focused on statistical methods for functional data classification and integer-valued time series. Email: amanda.fernandez-fontelo@hu-berlin.de

**Pascal. J. Kieslich** is a postdoctoral researcher at the Mannheim Centre for European Social Research (MZES) and the Experimental Psychology Lab, University of Mannheim, Germany. He received a master's degree and PhD in Psychology from the University of Mannheim. His research focuses on how people make decisions in different domains, and how decision processes can be traced over time. He is the lead developer of the mousetrap package for analyzing mouse-tracking data, and contributor to the mousetrap-web data collection framework. Email: kieslich@psychologie.uni-mannheim.de

**Felix Henninger** is a cognitive scientist and research software engineer, working at the Mannheim Centre for European Social Research (MZES) and the University of Koblenz-Landau Cognition Lab. His research investigates the mental processes underlying decisions and judgments, with a focus on risky choices, as well as tools for online experimentation and distributed data collection. He is the primary author of the mousetrap-web data collection framework, and contributor to the mousetrap analysis package. Email: mailbox@felixhenninger.com

**Frauke Kreuter** is a professor in the joint program in survey methodology at the University of Maryland, USA; professor of statistics and methodology at the University of Mannheim; and head of the Statistical Methods Research Department at the Institute for Employment Research in Nürnberg, Germany. She received her master's degree in sociology from the University of Mannheim, Germany, and her PhD in survey methodology from the University of Konstanz. Her research focuses on sampling and measurement errors in complex surveys. Email: fkreuter@umd.edu

**Sonja Greven** is a professor of statistics at the Humboldt-Universität zu Berlin, Germany. She received two Masters' degree in mathematics and biostatistics from RWTH Aachen University and the University of North Carolina at Chapel Hill, respectively, and a PhD in Statistics from Ludwig-Maximilians-Universität München. Her research concentrates on statistical methods for complex data such as functions, trajectories, images, densities or shapes. Email: sonja.greven@hu-berlin.de

# Data Availability

Data used in this article are available at the Institute for Employment Research (IAB). Up-to-date access information can be found here (https://www.iab.de/en/daten.aspx).

# Software Information

Statistical analysis has been performed using the free R 3.6.1 software.

# Supplementary material

The online supplementary materials and R codes are available at

https://github.com/PascalKieslich/mtdifficulty

**Table 1**: Computed mouse-tracking measures with the mousetrap package in R. All hover measures exclude a potential initial phase without mouse movement (which is reflected in the initiation time).

| Type | Measure | Definition |
|---|---|---|
| time | response time (RT) | time from page load until response submission |
|  | initiation time | the duration from page load until the first recorded cursor movement occurred |
| hovers | number of hovers | number of periods without movement exceeding a minimum duration threshold |
|  | overall duration of hovers | total time of all periods without movement exceeding a minimum duration threshold |
| distance | total distance | Euclidean distance travelled by the cursor |
| derivatives | maximum velocity | maximum movement velocity |
|  | maximum acceleration | maximum movement acceleration |
| flips | x-flips | number of changes in movement direction along horizontal axis |
|  | y-flips | number of changes in movement direction along vertical axis |

**Table 2**: Best predictive models and performance results for uncorrected and corrected full and response-time-only models. The threshold for the hovers was selected to optimize accuracy in the inner loop of the response-time-only cross-validation. Largest accuracy for each question indicated in bold.

*The same result was found for more than this threshold - here, the lowest threshold among the set of thresholds with the same accuracy is reported.

| question | manipulation | personalization | model | supervised learner | threshold hovers (ms) | accuracy | sensitivity | specificity |
|---|---|---|---|---|---|---|---|---|
| employment detail | concise vs. complex language | no | full | classification tree | 250 * | 0.6097 | 0.2951 | 0.8828 |
| | | | response-time-only | logit regression | | 0.6171 | 0.6358 | 0.6028 |
| | | baseline | full | tree-based gradient boosting | 2000 | **0.6587** | **0.5629** | **0.7416** |
| | | | response-time-only | classification tree | | 0.6407 | 0.6430 | 0.6376 |
| | | baseline and position | full | classification tree | 250* | 0.6498 | 0.7256 | 0.5772 |
| | | | response-time-only | classification tree | | 0.6480 | 0.5722 | 0.7065 |
| employee level | ordered vs. unordered layout | no | full | tree-based gradient boosting | 3000 | 0.5548 | 0.4695 | 0.6303 |
| | | | response-time-only | tree-based gradient boosting | | 0.5569 | 0.4420 | 0.6533 |
| | | baseline | full | neural network | 3000 | 0.5670 | 0.7015 | 0.4340 |
| | | | response-time-only | classification tree | | 0.5528 | 0.7168 | 0.3919 |
| | | baseline and position | full | tree-based gradient boosting | 3000 | **0.5909** | **0.5258** | **0.6527** |
| | | | response-time-only | neural network | | 0.5328 | 0.5498 | 0.5166 |
| education level | | no | full | support vector machines | 2000 | 0.5622 | 0.6042 | 0.5204 |
| | | | response-time-only | tree-based gradient boosting | | 0.5641 | 0.6502 | 0.4783 |
| | | baseline | full | neural network | 250 | 0.5805 | 0.7053 | 0.4436 |

| | | response-time-only | tree-based random forest | | 0.5639 | 0.5489 | 0.5831 |
| | baseline and position | full | tree-based random forest | 250 | **0.5895** | **0.6233** | **0.5512** |
| | | response-time-only | logit regression | | 0.4946 | 0.6117 | 0.4095 |

**Table 3**: Models' performance results for the best models and other full-model candidates that used the same personalization methods and thresholds for hovers as the corresponding best models for each target question.

| question | manipulation | personalization | supervised learner | accuracy | sensitivity | specificity |
|---|---|---|---|---|---|---|
| employment detail | concise vs. complex language | baseline | logit regression | 0.6298 | 0.6448 | 0.6169 |
| | | | classification tree | 0.6189 | 0.5836 | 0.6569 |
| | | | tree-based random forest | 0.6134 | 0.5516 | 0.6664 |
| | | | tree-based gradient boosting | **0.6587** | **0.5629** | **0.7416** |
| | | | support vector machines | 0.6153 | 0.5755 | 0.6498 |
| | | | neural network | 0.5628 | 0.6866 | 0.4513 |
| employee level | ordered vs. unordered layout | baseline and position | logit regression | 0.4751 | 0.3062 | 0.6680 |
| | | | classification tree | 0.5709 | 0.5275 | 0.6184 |
| | | | tree-based random forest | 0.5748 | 0.5286 | 0.6078 |
| | | | tree-based gradient boosting | **0.5909** | **0.5258** | **0.6527** |
| | | | support vector machines | 0.5709 | 0.2823 | 0.8435 |
| | | | neural network | 0.5190 | 0.6677 | 0.3780 |
| education level | | baseline and position | logit regression | 0.4820 | 0.5666 | 0.4259 |
| | | | classification tree | 0.5765 | 0.7013 | 0.4673 |
| | | | tree-based random forest | **0.5895** | **0.6233** | **0.5512** |
| | | | tree-based gradient boosting | 0.5748 | 0.6683 | 0.4851 |
| | | | support vector machines | 0.4634 | 0.4145 | 0.5644 |
| | | | neural network | 0.5222 | 0.5927 | 0.4249 |

**Figure 1:** Distribution of the uncorrected mouse-tracking measures separately for each question (education level, employee level, and employment detail from left to right) and difficulty condition (blue = easy, red = difficult). The smoothed kernel density estimates are presented. Axis limits are set so that for each question and measure > 95 % of the values are displayed.

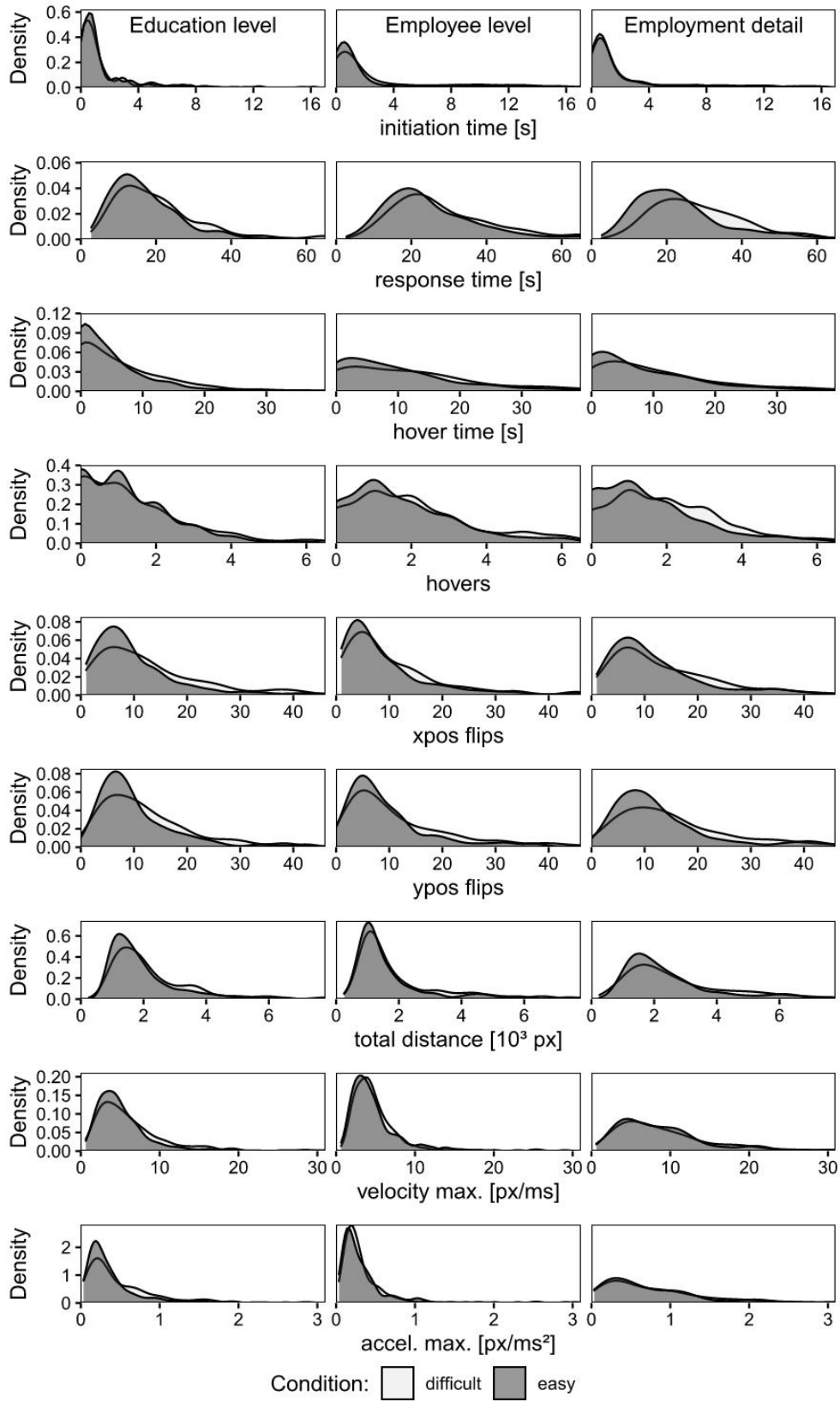

**Figure 2:** Measures' importance based on permutational methods. The impact of a measure is evaluated through the model's accuracy reduction when this measure is randomly permuted. The more negative the reduction, the more impact the measure has in the model.

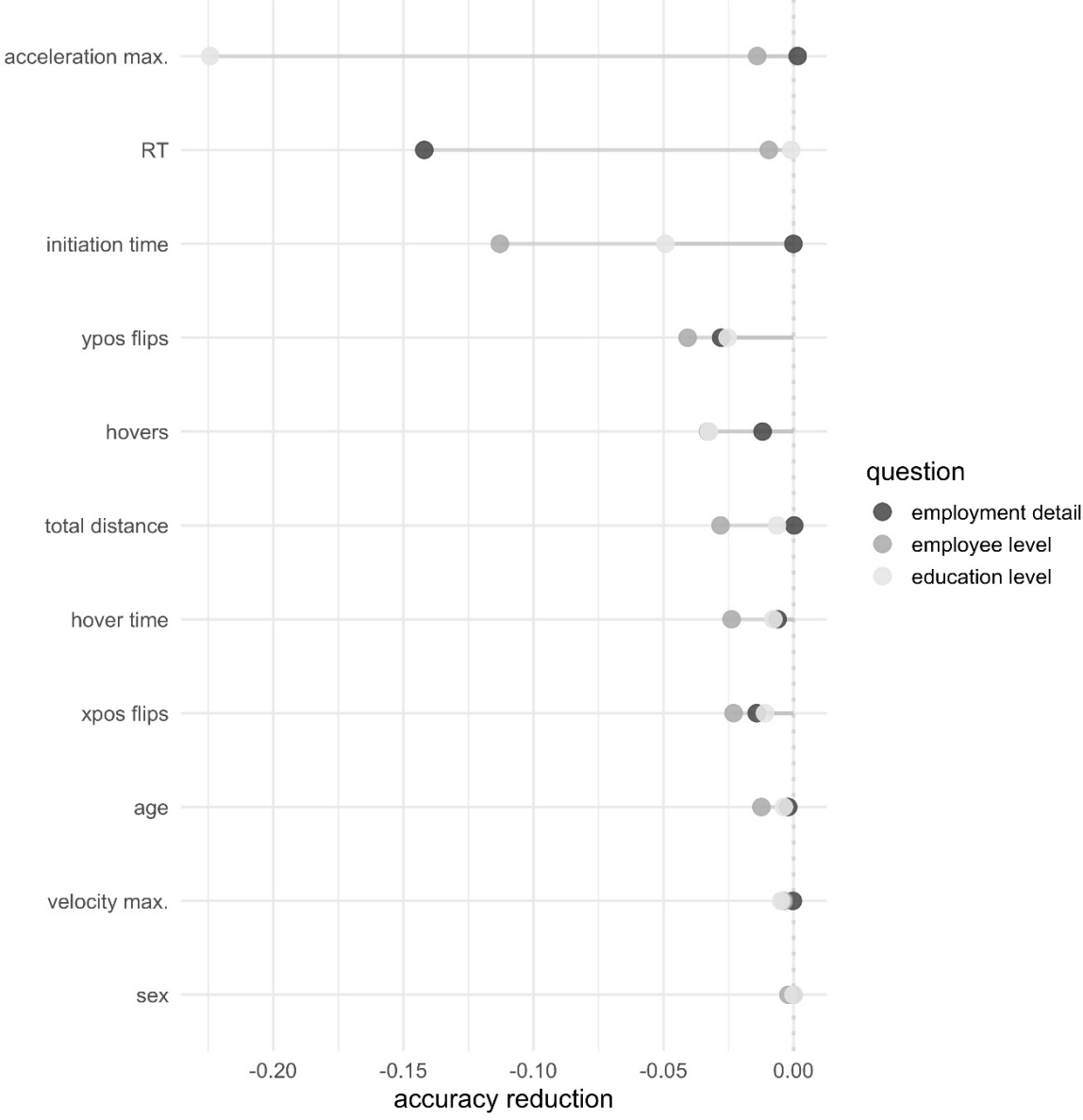